\documentclass{aa}
\usepackage{epsf}
\usepackage{times}
\newcommand{\etal}{{et al}\/.}
\begin{document}
\title{Low-frequency constraints on the spectra of the lobes of
  the microquasar GRS 1758$-$258}
\titlerunning{Lobes of GRS 1758$-$258}
\author{M. J. Hardcastle}
\institute{Department of Physics, University of Bristol, Tyndall
Avenue, Bristol BS8 1TL, UK \and School of Physics, Astronomy and Mathematics, University of
  Hertfordshire, College Lane, Hatfield AL10 9AB, UK (mjh@star.herts.ac.uk)}
\date{This version submitted 21st September 2004: accepted 22nd December 2004}

\abstract{I present low-frequency observations of the lobes of the
  microquasar GRS 1758$-$258, which, together with earlier radio
  observations, provide the best constraints yet on the lobe spectra,
  and show that they are consistent with the electron energy indices
  expected from standard shock acceleration models. There is no
  statistically significant evidence for synchrotron ageing along the
  length of the lobes, which is consistent with expectation if the
  magnetic field strength is close to the minimum-energy value. If the
  environment of the lobes has typical ISM properties, the lobes will
  dissipate on timescales of thousands of years or less, long before
  synchrotron ageing can become significant.  }

\maketitle

\section{Introduction}

Galactic `microquasar' sources, driven by relativistic outflows from
close to the accretion discs of neutron star or black hole binaries
(Mirabel \& Rodr\'\i guez 1999) are of great interest, not just for
the insights they provide into the properties of accreting binary
systems but as analogues of the much more powerful and distant
extragalactic radio sources, radio galaxies and radio-loud quasars.
About 10\% of known X-ray binary systems have detected radio emission
which may be associated with jets, and a significant fraction of these
show, or have at some stage shown, clear jet-related extended radio
emission (Mirabel \& Rodr\'\i guez). However, there is one crucial
difference between microquasars and their extragalactic counterparts.
Essentially all powerful AGN-related radio sources, when observed at
sufficient dynamic range, show extended `lobes' of radio emission,
which are thought to be the repository for a significant fraction of
the jet's overall luminosity integrated over the radio source's lifetime (e.g.
Scheuer 1974) and which very often dominate the radio luminosity of
the source by many orders of magnitude. By contrast, there are only
two known microquasars that have so far been found to exhibit
persistent lobe-like radio emission: the two Galactic-centre black
hole candidates 1E 1740.7$-$2942 (Mirabel \etal\ 1992) and GRS
1758$-$258, the subject of this paper. Heinz (2002) has argued that
this difference arises because of the (relatively) low-density
environments generally inhabited by microquasars, which mean that any
lobes formed have short lifetimes against adiabatic losses; this might
suggest that there is something special about the environments of the
sources that show lobes. However, because of the very low radio flux
from the lobed objects, it has been difficult so far to make
quantitative statements about their dynamics. In this paper I present
new observations that allow more definite statements to be made about
GRS 1758$-$258.

GRS 1758$-$258 is a bright persistent hard X-ray source lying 5
degrees away from the Galactic centre, originally discovered with the
SIGMA instrument on {\it Granat} (Sunyaev \etal\ 1991).
(Throughout this paper I shall follow previous work in adopting a
distance of 8.5 kpc to the source.) Because of the high extinction
towards GRS 1758$-$258 in the optical, searches for the companion star
have been carried out in the infrared (e.g. Mart\'\i\ \etal\ 1998,
Rothstein \etal\ 2002): the most likely counterpart, based on radio
and X-ray positions, is a K-III-type giant star. The associated
extended radio structure was originally discovered by Rodr\'\i guez,
Mirabel \& Mart\'\i\ (1992) and subsequently imaged in more detail at
two frequencies by Mart\'\i\ \etal\ (2002: hereafter M02). M02 used
the fact that the spectral indices between 5 and 8.4 GHz were
consistent with being steep ($\alpha > 0$, where $\alpha$ is defined
throughout the paper in the sense that flux $S \propto \nu^{-\alpha}$)
to argue that the lobe emission was definitely optically thin
synchrotron radiation. The new radio observations presented in this
paper, combined with the data of discussed by M02, provide much
tighter constraints on the radio spectral properties of the lobes, and
allow me to discuss their implications for the dynamics of the radio
source.

Throughout the paper errors quoted are $1\sigma$, unless otherwise stated.

\section{Observations and analysis}

\begin{figure*}
\epsfxsize 5.9cm
\epsfbox{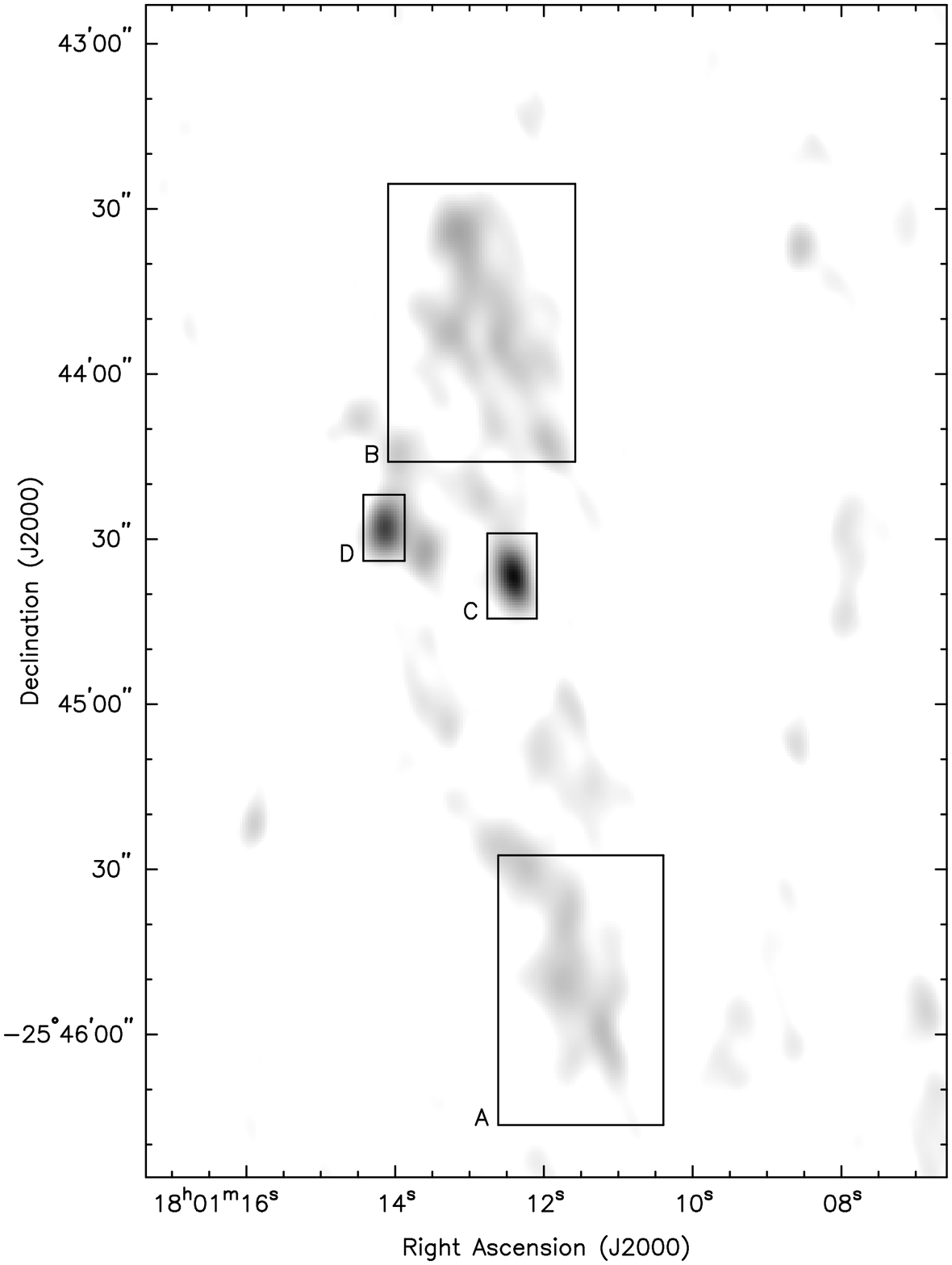}
\epsfxsize 5.9cm
\epsfbox{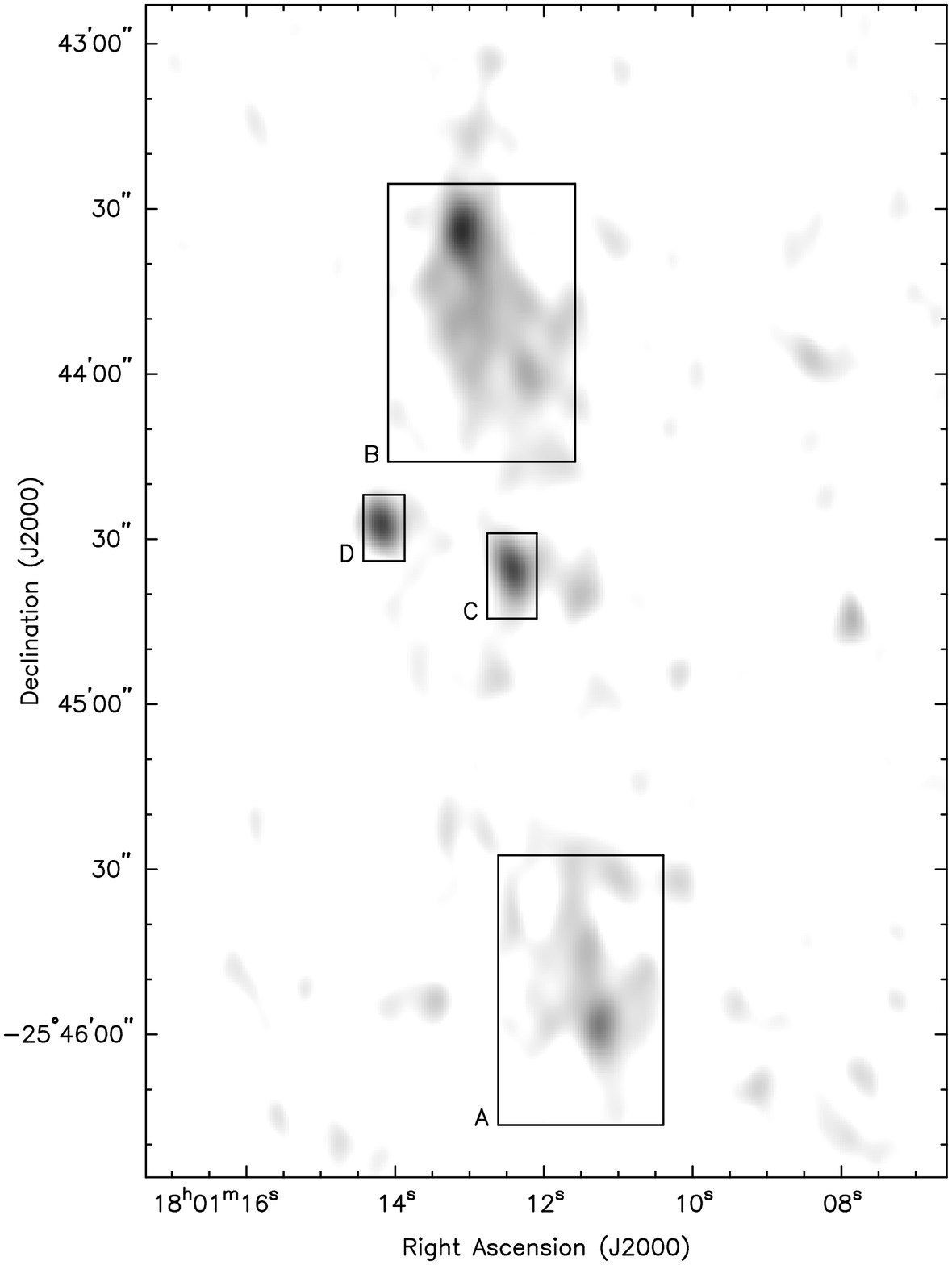}
\epsfxsize 5.9cm
\epsfbox{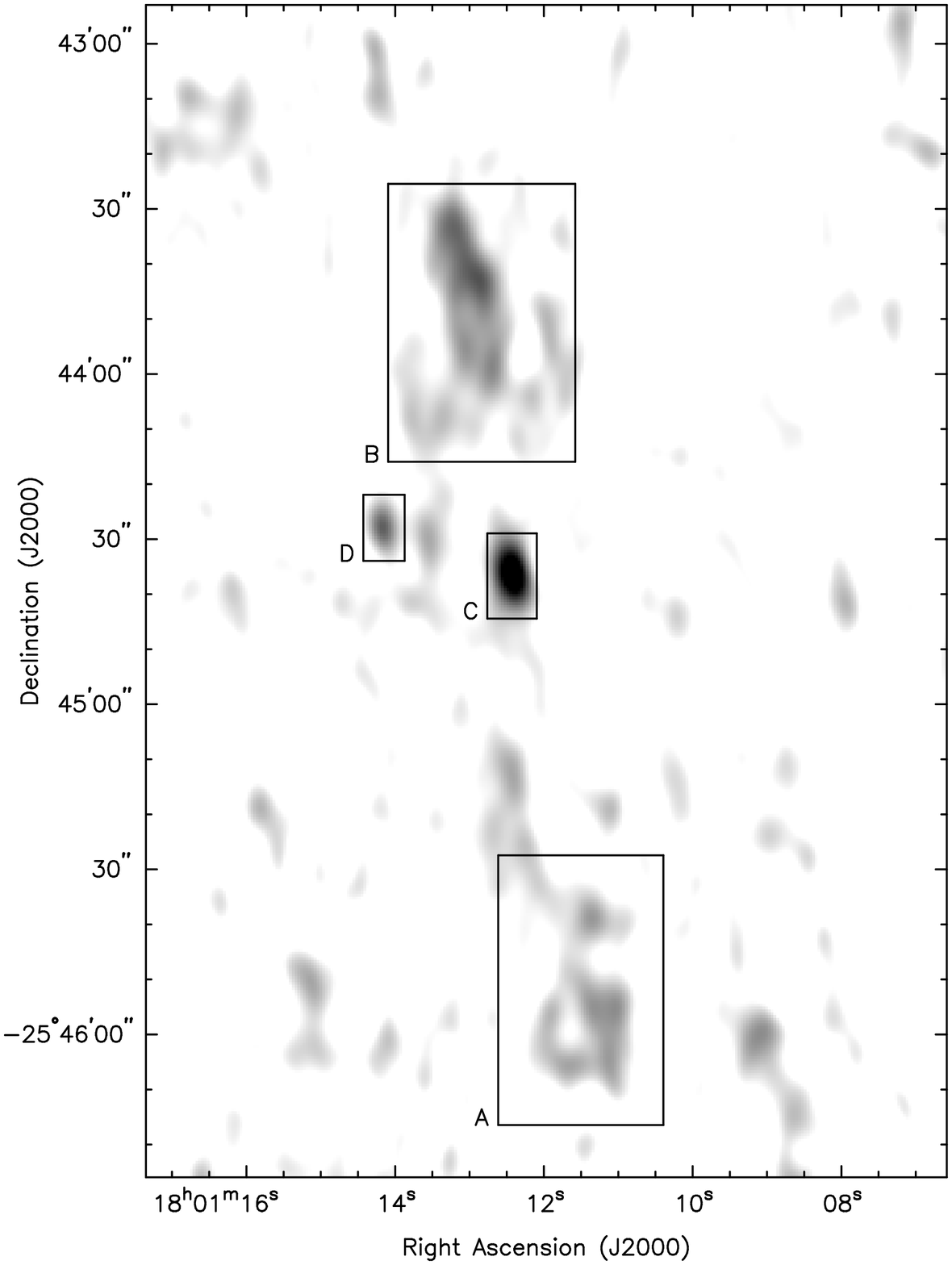}
\caption{Greyscale images of the GRS 1758$-$258 lobes at the three
  frequencies used in the paper: 1.4 GHz (left), 4.9 GHz (centre) and
  8.4 GHz (right). All three images have the same resolution, $9.2
  \times 5.8$ arcsec. The greyscale ranges from the $2\sigma$ level of
  the off-source noise to 0.5 mJy beam$^{-1}$ (1.4 GHz), 0.2 mJy
  beam$^{-1}$ (4.9 GHz) and 0.15 mJy beam$^{-1}$ (8.4 GHz). Boxes show
  the extraction regions used for the four source components discussed
  in the paper. Component C is identified with the X-ray source.}
\label{images}
\end{figure*}

GRS 1758$-$258 was observed twice with the NRAO Very Large Array (VLA)
in B configuration, using two standard observing frequencies (1385.1
and 1464.9 MHz) with a bandwidth of 50 MHz at each frequency. The
first observation, made on 2002 August 11, was badly compromised by
radio frequency interference, and accordingly I re-observed the source
with the same observational parameters on 2004 Jan 13. The total
usable on-source integration time was about 3h. 3C\,286 was used as
the primary flux calibrator and the phase was calibrated using
observations of the nearby source 1751$-$253 every 15 minutes. The
data were calibrated in the software package AIPS in the standard
manner, and the task IMAGR was used to produce images of the source at
an effective frequency of 1425 MHz. Because there was no evidence for
significant changes in the structure of the source over the 16-month
interval between observations (see below for more discussion of this
point) I combined the two observations to make a final image. The
off-source noise on this image (35 $\mu$Jy beam$^{-1}$) was
significantly above the expected thermal level, but was sufficient for
a significant detection of all the previously known components of the
radio source (Fig. \ref{images}).

In order to make reliable multi-frequency measurements, I retrieved
the data discussed by M02 from the VLA archives.
This consists of 9 observations at C and X bands in the C
configuration of the VLA taken between 1997 August 03 and 1997 August
24 as part of a multi-wavelength variability campaign (see Lin \etal\
2000 for the observational details). Following M02, I
calibrated each of the C and X-band datasets separately and
concatenated them to make images at effective frequencies of 4860 and
8460 MHz. The $uv$-plane coverage of my new L-band data is well
matched to that of the 1997 C-band data, but the X-band data are
obviously not as well matched; at the time of writing, there are no
publicly accessible D-configuration X-band observations of useful
length that could be used to provide short baselines at 8 GHz.
Accordingly, I elected to image all three datasets with a shortest
baseline of 0.7 k$\lambda$, matched to the shortest baseline of the
8-GHz dataset, to ensure that the same largest angular size was
sampled by each. As this short-baseline cutoff has significant effects
on spatial scales above 2.5 arcmin, it should not affect
the images produced. The weighting of the $uv$ plane was tapered in
imaging the X-band data to produce a similar beam area, and primary
beam correction was applied to both the C- and X-band images using the
AIPS task PBCOR.

I then measured flux densities from four regions of the source,
corresponding to the components denoted VLA A, B, C and D by
Rodr\'{\i}guez \etal\ (1992). A is the south lobe, B the north lobe, C
is coincident with the {\it Chandra} position of the X-ray source
(Heindl \& Smith 2002) and seems likely to be analogous to the `core'
of an extragalactic radio source (see below) while D is thought to be
an unrelated object. Flux densities were measured in rectangles around
each component using the AIPS verb IMSTAT. The regions used to extract
the fluxes are shown in Fig.\ \ref{images}, and the flux densities
listed in Table \ref{fluxes}. Errors on the flux densities were
determined from the off-source noise in each image in the standard
way. Note that the flux densities for C and X bands are systematically
higher than the values tabulated by M02: this
seems likely to be because I have used larger measurement regions, in
an attempt to measure all the lobe-related flux.

\begin{table}
\caption{Flux densities measured for the different components of GRS~1758$-$358}\label{fluxes}
\begin{tabular}{lrrr}
\hline
Component&\multicolumn{3}{c}{Flux density (mJy)}\\
&(1.4 GHz)&(4.9 GHz)&(8.4 GHz)\\
\hline
A&$1.51 \pm 0.23$&$0.80 \pm 0.09$&$0.59 \pm 0.10$\\
B&$2.35 \pm 0.25$&$1.25 \pm 0.09$&$1.01 \pm 0.11$\\
(B north)&$1.04 \pm 0.18$&$0.68 \pm 0.07$&$0.49 \pm 0.07$\\
(B south)&$1.31 \pm 0.18$&$0.56 \pm 0.07$&$0.52 \pm 0.07$\\
C&$0.52 \pm 0.05$&$0.19 \pm 0.03$&$0.23 \pm 0.03$\\
D&$0.37 \pm 0.05$&$0.14 \pm 0.02$&$0.08 \pm 0.03$\\
\hline
\end{tabular}
\end{table}

\begin{figure}
\epsfxsize 8.7cm
\epsfbox{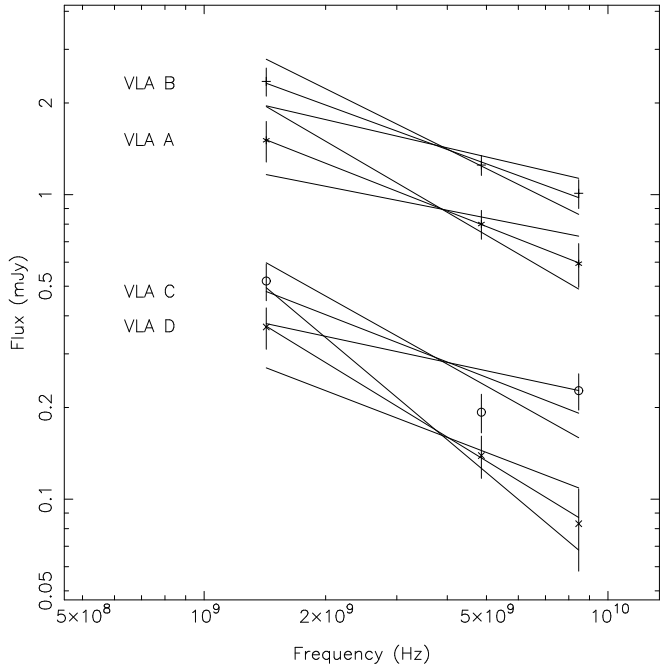}
\caption{The flux densities and best power-law fits of the
  components of GRS~1758$-$358. The solid lines show the best-fitting
  lines and the 90\% confidence limits on the slope.}
\label{powerlaw}
\end{figure}

The flux densities were then fitted with a power law in frequency
using standard least-squares fitting (Fig.\ \ref{powerlaw}). The lobe
regions (A and B) are both well fitted with a power law, with spectral
indices $\alpha$ of $0.52 \pm 0.12$ and $0.49 \pm 0.08$ respectively,
in good agreement with the expectation from synchrotron radiation from
a population of particles accelerated at a strong shock ($\alpha =
0.5$). There is no evidence for any high-frequency spectral
steepening. The central source, C, is not well fitted with a power-law
model because of the upturn in the flux density between 5 and 8 GHz,
but this may well be due to the effects of variability (see section
\ref{variability} below). Component D is well fitted with a power law
of slope $0.81 \pm 0.15$, which would be a plausible value for a
background radio galaxy.

Finally, in view of the differences between the images in L and C
band, suggesting that the source is more strongly edge-brightened at
higher frequencies, I measured fluxes for two sub-regions of component
B, obtained by dividing the region shown in Fig.\ \ref{images} into
two equal halves. The fitted spectral indices for these two regions
are $0.40 \pm 0.12$ and $0.55 \pm 0.10$. There is thus little
statistically significant evidence for spectral index variation along
the length of the northern lobe.

\section{Discussion}

\subsection{Source variability}
\label{variability}
An important question is whether the radio source components have
varied on timescales of years, and, in particular, whether there could
have been significant variability between 1997 and 2002/2004, the
respective epochs of the C and X band measurements and my new L-band
measurements. Strong variability would invalidate the spectral index
results presented above. The images show no evidence for {\it motion}
of the lobes, and this is also the case for the 1992 and 1997 C-band
images (M02). There is no detectable variability in the 1.4-GHz lobe
fluxes (or the fluxes of the other components) between 2002 and 2004,
but the errors are comparatively large because the 2002 data are poor:
M02 comment that they were unable to determine whether there was
variability between 1992 and 1997. We know that component C {\it is}
variable on comparatively short timescales (Rodr\'\i guez \etal\ 1992)
although there was little evidence for strong variability in the
monitoring campaign of Lin \etal\ (2000): if the L-band flux in 1997
was consistent with the flat spectral index they measured between 4.9
and 8.4 GHz, however, then it must have increased by a factor 2--3
between 1997 and 2002/4. Variability would certainly explain the
peculiar spectrum of this component. Component D appears to have
varied between 1992 and 1997 (Lin \etal ), which is perhaps surprising
in view of its steep spectrum. In the absence of any detected
variability in the lobes, I will assume in what follows that the
measured spectra for the lobes are valid. Data (at the time of
writing, still proprietary) exist in the VLA archive that should allow
a definitive answer to this question.

\subsection{Synchrotron lobes}

The agreement between the measured lobe spectral indices and the
standard spectral index for shock-accelerated synchrotron emission is
striking. It allows us to assume a synchrotron model with a good deal
of confidence, but it also suggests strongly that the lobes are
generated in a manner similar to that of powerful extragalactic radio
sources. A low-frequency spectral index of 0.5 is often observed in
the hotspots of such objects (e.g. Meisenheimer \etal\ 1989).

If we adopt a synchrotron model, then we can obtain revised versions
of the standard derived quantities. I model the electron spectrum
as a power law with energy spectral index of 2, corresponding to
$\alpha = 0.5$, and assume that the electron energies extend down to
$\gamma = 1$ and up to $\gamma = 10^5$, and that there are no
relativistic protons. The derived energy density is only
  logarithmically dependent on the assumed low and high-energy cutoffs
  in the electron spectrum. The N lobe (B), which I model
as an ellipsoid with major axis 55 arcsec and minor axis 25 arcsec,
then has a minimum-energy magnetic field strength of 2.7 nT; the minimum
lobe energy density is $7 \times 10^{-12}$ J m$^{-3}$, and the minimum
pressure in the lobe is $2 \times 10^{-12}$ Pa. The minimum total
energy stored in the lobe is $2 \times 10^{38}$ J, comparable to the
earlier estimate by Rodr{\'\i}guez \etal\ (1992).

If we assume that the dominant loss mechanism for the electrons is
synchrotron radiation\footnote{The magnetic field energy density,
assuming minimum-energy conditions, appears to dominate over the
photon energy density from the central star and X-ray source at the
distance ($\sim 2$ pc) of the lobes. An estimate of the overall photon
energy density near the Galactic centre is hard to obtain.}, then
standard results (e.g. Leahy 1991), assuming that particle
acceleration initially produces an electron energy spectrum that
extends to well above the energies being observed at GHz frequencies,
show that, for the minimum-energy magnetic field strength, the effects
of ageing will not become apparent at 8 GHz until $4 \times 10^6$
years, after particles in that field were last accelerated. (The
similarity between this and the `radiative loss' timescale quoted by
Heindl \& Smith (2002) is coincidental.) Since these timescales are
very much longer than the characteristic timescales of the jet source,
we cannot expect to be able to use spectral ageing at GHz frequencies
as a probe of lobe dynamics, as is often done for extragalactic radio
sources (e.g. Alexander \& Leahy 1987) unless the magnetic fields are
very much stronger than the minimum energy estimate. In extragalactic
sources, there is independent evidence from inverse-Compton emission
that the field strengths are close to the minimum-energy values (e.g.
Hardcastle \etal\ 2002) but we have no such independent constraints in
microquasars, nor are they likely to be easy to obtain. However, the
long synchrotron lifetimes are consistent with the observed lack of
any significant spectral gradient along the northern lobe.

\subsection{Dynamics and expansion}

I begin this section by noting that it is not clear whether there is
continuing energy supply to the lobes of GRS 1758$-$258. The observed
radio variability in component C suggests ongoing jet activity, but
this may be at a very low level. The long lifetimes of the lobes
derived above mean that we cannot rule out the possibility that they
are the result of an outburst of efficient jet activity at some time
in the past (to be determined, but $\la 10^6$ years ago). As Heindl \&
Smith (2002) point out, the entire stored energy of the lobes can be
supplied in a relatively short time (a few years) if the accretion
luminosity is tapped efficiently. For this reason, I consider below
models of the lobes in which the energy supply from the jet can be
neglected and the dynamics of the lobes are driven by their internal
energy. In this case the evolution of the radio flux density of the
lobes is determined by adiabatic expansion; we expect flux density to
go as $r^{2+4\alpha}$ (e.g. Longair, Ryle \& Scheuer 1973).

The minimum pressure in the lobes is very similar to that of the dense
molecular material found near the Galactic centre (e.g. Martin \etal\
2003), so that the lobes could be confined (and therefore not evolving) if they lay inside a
molecular cloud. A molecular cloud is known to lie in the direction of
the source (Lin \etal\ 2000) but the comparatively low absorbing
column ($2 \times 10^{22}$ cm$^{-2}$: Lin \etal\ and references
therein) would require the source to lie relatively close to the
surface of the cloud. This is by no means impossible, and the
similarity between the minimum pressure and the available external
pressure is interesting. However, if the external pressure is lower
than this, the lobes cannot be confined. In discussing this case we
take (for simplicity, rather than out of any belief that these
conditions obtain near the Galactic centre) a fiducial external
density $n_0 \approx 1$ cm$^{-3}$. A lower limit on the expansion speed
at the present time, using ram pressure balance, is then around
$35(n/n_0)^{-1/2}$ km s$^{-1}$.  Assuming expansion according to a
Sedov law, this would give a timescale for a significant drop in flux
density, say by a factor $\sim 2$, of $10^4 (n/n_0)^{1/2}$ years --- less if
the lobe pressure exceeds the minimum value. Adiabatic expansion, not
synchrotron losses, would determine the timescale on which the lobe can
be detected in this scenario (and this is true for any plausible
magnetic field strength in the lobes). This would again be in contrast to the
situation for extragalactic sources, where the minimum-energy
pressures in lobes routinely fall {\it below} the external thermal pressure
(e.g. Hardcastle \& Worrall 2000), and where approximate pressure
balance (with moderate departures from the minimum-energy condition)
can be inferred based on X-ray emission (e.g. Croston \etal\ 2004).

The lobes do not appear to be moving relativistically; M02 estimate an
expansion speed $<0.3c$. If we assume that the lack of detectable
variability between 2002 and 2004 implies that the lobes do not vary,
the synchrotron model allows us to place constraints on any adiabatic
expansion that may be taking place: a radial expansion of 10\% on a
timescale of years (corresponding to expansion speeds of the order of
$0.1c$) would give rise to a decrease in flux density of a factor 1.5,
which should have been detectable even in the existing poor-quality
two-epoch data. We can conclude from this only that the lobes are not
grossly overpressured (by factors $\sim 10^7$, for $n = n_0$)
with respect to the ambient medium. Longer temporal
baselines and more accurate flux density monitoring would provide much
stronger constraints.

\section{Conclusions}

To summarize, the main conclusions that can be drawn from the accurate
measurement of the spectral index of the lobes of GRS 1758$-$258
presented here are as follows:

\begin{enumerate}
\item The good agreement between the spectral indices and the expected
  value for synchrotron emission from shock-accelerated electrons
  strongly reinforces earlier conclusions that the radio emission
  process is synchrotron radiation.
\item The lack of significant difference between the spectral indices
  at different regions in the N lobe suggests that synchrotron losses
  are not important at GHz frequencies: this is consistent with
  expectation if the magnetic field strengths are close to the
  minimum-energy values and the source has an age $\la 10^6$ years.
\item The lobe minimum pressures on a synchrotron model are comparable
  to the internal pressures in Galactic centre molecular clouds, and
  so the lobes could be confined by such a system. If the lobes lie in
  a less dense environment they will have a lifetime of the
  order of at most thousands of years before adiabatic losses render them
  undetectable, assuming there is no further energy supply from the
  jet.
\end{enumerate}

\begin{acknowledgements}
I thank the Royal Society for a Research Fellowship, and Christian
Kaiser for originally pointing out to me the existence of double-lobed
microquasars. J. Mart\'\i\ kindly supplied me with a version of the
map of Mart\'\i\ \etal\ (2002) in support of this work. I am grateful
to John Benson for help in retrieving the archival VLA data. I thank
the anonymous referee for comments that allowed me to improve the
paper significantly.

The National Radio Astronomy Observatory is a facility of the National
Science Foundation operated under cooperative agreement by Associated
Universities, Inc.
\end{acknowledgements}

\end{document}